\documentclass[final,3p,times,twocolumn]{elsarticle}
\journal{ }

\usepackage{pdfpages}

\usepackage{xcolor}
\usepackage{amsmath}
\usepackage{amssymb}

\usepackage{mathtools}

\DeclarePairedDelimiter\ket{\lvert}{\rangle}
\DeclarePairedDelimiterX\braket[2]{\langle}{\rangle}{#1\,\delimsize\vert\,\mathopen{}#2}

\newcommand{\abs}[1]{\left\lvert #1\right\rvert}
\newcommand{\dif}{\,\mathrm{d}}

\begin{document}

\begin{frontmatter}

\title{Terahertz-driven four-wave mixing at glass surfaces: Probing vibrational resonances and structural regimes}

\author[inst1]{Mathias H. Kristensen \corref{cor1} \fnref{fn1}}
\ead{mhkr@icloud.com}

\fntext[fn1]{Currently at CROMA UMR 5130 (Université Savoie Mont Blanc, Université Grenoble Alpes, Grenoble INP, and CNRS), 73370 Le Bourget du Lac, France.}

\cortext[cor1]{Corresponding authors.}

\affiliation[inst1]{organization={LOMA UMR 5798, University of Bordeaux and CNRS},
            city={33405 Talence},
            country={France}}

\author[inst1]{Jérôme Degert}
\author[inst1]{Theo Guillaume}
\author[inst1]{Emmanuel Abraham}
\author[inst1]{Laetitia Dalstein}
\author[inst1]{Eric Freysz\corref{cor1}}
\ead{eric.freysz@u-bordeaux.fr}

\begin{abstract}
Disordered materials such as glasses exhibit complex structural dynamics that are challenging to probe with conventional spectroscopies.
We demonstrate that terahertz-driven four-wave mixing (FWM) at glass surfaces provides direct access to low-frequency vibrational modes and structural evolution in amorphous solids.
Applied to a compositional series of PbO--silicate glasses (20--54~mol\% PbO), this technique resolves distinct contributions from collective Boson-peak excitations and Pb--O / Si--O network stretching modes, and tracks their systematic evolution across structurally distinct compositional regimes.
The dominant vibrational frequency blueshifts with PbO content, reflecting the progressive evolution of the Pb$^{2+}$ network role from silicate-modifier to ward network-former.
A pronounced enhancement of the FWM signal near 44~mol\% PbO coincides with the emergence of medium-range Pb--Pb correlations, while in-plane--to--out-of-plane FWM intensity ratio ($I_{\rm SS}/I_{\rm PS}$) tracks $\chi^{(3)}$ tensor anisotropy tied to Pb$^{2+}$ lone-pair spatial correlations.
The non-monotonic peak in both observables at 44~mol\% PbO -- a composition where NMR finds no change in local Pb--O coordination and Pb--O--Pb free-oxide linkages are negligible -- provides direct evidence that a collective lone-pair reorganization occurs in the medium-range structure independently of nearest-neighbor bonding.
These results establish terahertz-driven FWM as a bulk-sensitive, near-surface depth-confined ($\sim$50~nm) nonlinear spectroscopy sensitive to vibrational and electronic structural fingerprints inaccessible to linear infrared, Raman, and terahertz  time-domain probes.
\end{abstract}

\end{frontmatter}

Nonlinear light--matter interactions driven by terahertz (THz) fields provide direct access to ultrafast dynamics and low-energy excitations in a wide range of systems, from solids to liquids, gases, and complex interfaces, making them invaluable for studying structural, electronic, and chemical processes \cite{Nicoletti2016nonlinearTHz, Lu2024nonlinearTHz}. 
Intrinsic nonlinear responses at THz frequencies have also been observed \cite{Lu2024nonlinearTHz,Selz2025}.
In particular, strong-field THz spectroscopy enables access to regimes where conventional optical probes are either insensitive or perturbative, opening new avenues for studying phonons \cite{Dalstein2024}, polaritons \cite{Paparo2025}, and other collective modes in solids \cite{Frenzel2023,Vaswani2021,Grasset2022}.
This capability is especially relevant in disordered systems such as glasses, where the interplay between structural heterogeneity and vibrational dynamics gives rise to complex physical behavior.

The definition of amorphous materials has broadened in recent years to include not only conventional oxide glasses but also hybrid and metal–organic framework (MOF) glasses with remarkable compositional diversity and emergent functionality \cite{Bennett2024}. A key structural feature is medium-range order (MRO) -- correlations extending beyond nearest-neighbor bonding, typically spanning 5–20~\AA{} and including motifs such as edge-sharing polyhedra and ring structures \cite{Sorensen2020}.
Although not directly accessible by diffraction, growing evidence shows that MRO correlates with glass properties including thermal and mechanical properties \cite{Shi2023,Sorensen2020,Sorensen2022,Christensen2023, Baldi2010, Gelin2016}, as well as the structural relaxation dynamics associated with the glass transition \cite{Shintani2008}.

A distinctive manifestation of these complex dynamics is the Boson peak -- an excess in the vibrational density of states over the Debye prediction -- which is universally observed in glasses yet remains enigmatic in origin. The Boson peak and related low-frequency collective modes have been extensively studied by linear spectroscopies such as THz time-domain \cite{Wada2024,Sorensen2022,Ando2021,Tostanoski2023}, infrared (IR) \cite{Radica2024,Gautam2021}, and Raman spectroscopy \cite{Radica2024,Drewitt2022,Kacem2017,Yadav2015}, while the static structure is well characterized by diffraction, extended X-ray absorption fine structure (EXAFS), molecular dynamics (MD), and nuclear magnetic resonance (NMR) studies \cite{Drewitt2022,Alderman2021,Rybicki2001,Takaishi2005,Sen2024}. 

Despite decades of investigation and diverse models invoking density fluctuations, elastic heterogeneities, or generic disorder \cite{Graebner1986, Elliott1992,Sokolov1999,Lubchenko2003,Zhang2017}, no universal consensus has emerged on the microscopic origin of the Boson peak.
A recent unified theory by Baggioli and Zaccone \cite{Baggioli2020} reframes the Boson peak in terms of nonpropagating, disorder-induced vibrational modes -- so-called diffusons -- that arise near the Ioffe-Regel limit, linking vibrational anomalies to structural disorder and damping. These concepts provide a modern foundation for interpreting low-frequency dynamics in glasses. Recent experimental and theoretical work provides additional evidence linking these low-frequency modes to medium-range structural motifs and correlated clusters \cite{GonzalezJimenez2023, Kyotani2025}. In particular, spatially coupled, quasi-localized nonphononic modes -- whose frequency and intensity depend sensitively on disorder and glass preparation -- have emerged as key contributors to these vibrational anomalies \cite{Moriel2024}.

Although linear spectroscopies such as Raman and inelastic neutron scattering have been instrumental in characterizing vibrational features of glasses, their ability to resolve structure--vibration relationships is limited, especially in the low-frequency regime where Boson modes emerge.
Recent simulation studies have underscored this point, showing that interatomic potentials vary considerably in their ability to reproduce experimental vibrational density of states and thermal properties, despite yielding similar structural features \cite{AkirmakYamac2025}.

These limitations highlights the need for nonlinear spectroscopic approaches that are sensitive to local structure and vibrational coupling -- particularly in disordered systems, where vibrational coherence and damping reflect complex structural motifs. 
In this context, coherent four-wave mixing (FWM) techniques based on the third-order susceptibility $\chi^{(3)}$ provide direct and selective access to vibrational dynamics -- including those with hyper-Raman-like selection rules -- that are inaccessible or extremely weak in conventional linear and spontaneous hyper-Raman ($\chi^{(5)}$) spectroscopies \cite{Clerici2013,Noskovicova2024,McDonnell2024}.
While THz-FWM has been demonstrated in bulk transmission geometries in crystalline diamond \cite{Clerici2013} and very recently in fluorides \cite{Noskovicova2024}, two aspects remain unexplored: its applicability to amorphous, topologically disordered networks, and its use in a reflection geometry, where the process is naturally confined to a thin near-surface layer by phase mismatch (as previously shown for silicon \cite{Dalstein2024}).
Here we combine both: we apply THz-driven FWM to amorphous systems for the first time, in a depth-confined reflection geometry, demonstrating its power to elucidate the interplay between network topology, disorder, and collective vibrational dynamics.

\begin{figure}[ht]
\centering
\includegraphics[width=\linewidth]{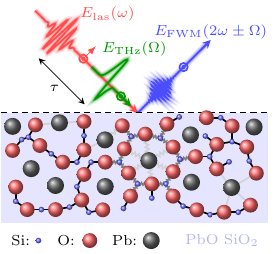}
\caption{
\textbf{THz-driven FWM at a lead oxide silicate glass surface.}
The incident THz electric field $E_\text{THz}(\Omega)$ and optical laser field $E_\text{las}(\omega)$ interact within the near-surface region to generate a FWM signal $E_\text{FWM}(2\omega \pm \Omega)$ at the corresponding Stokes (difference) and anti-Stokes (sum) frequencies. Spring-shaped bonds represent localized vibrational modes within the disordered lattice contributing to the nonlinear response. The delay time $\tau$ between the THz and optical pulses is indicated, enabling temporal control over the interaction.
}
\label{fig:FWM-concept}
\end{figure}

In this paper, we demonstrate THz-driven FWM at the surfaces of amorphous lead oxide silicate glasses (PbO content: 20--54~mol\%), as conceptually illustrated in Figure~\ref{fig:FWM-concept}. This nonlinear spectroscopic approach probes the vibrational landscape from collective low-frequency Boson modes to localized Pb--O stretching and Si--O network deformations. 
The FWM signal, oscillating at $2\omega \pm \Omega$, is centered around the surface second-harmonic (SHG) frequency but is dramatically enhanced by THz-driven coupling to vibrational modes \cite{Noskovicova2024}, analogously to phonon-enhanced sum-frequency generation in crystalline systems \cite{Mueller2026}.
The resulting signals are broadband and notably stronger than SHG alone, despite originating from the top $\sim$50~nm (Eq. \eqref{eq:interactionlength}) of the bulk.
Notably, the FWM emission is spectrally shifted toward the Stokes side (at $2\omega - \Omega$), with no corresponding anti-Stokes emission detected, consistent with resonant coupling to low-frequency modes.
By analyzing the spectral and polarization characteristics of the emission, we identify signatures of Boson modes and their nonlinear interactions, supported by the polarization dependence expected for a third-order FWM process.

Prior IR and Raman studies of PbO-silicate glasses \cite{Feller2010, Zagrai2024, Kacem2017} have established the key structural evolution with lead content: the progressive suppression of the silica Boson peak near 2.1~THz, the emergence of Pb--O stretching modes at 3 and 4.2~THz, and the systematic depolymerization of the Si--O network above $\sim$50~mol\% PbO. 
THz time-domain spectroscopy and hyper-Raman scattering further confirm the persistence of low-frequency collective modes in these glasses \cite{Wada2024, Hehlen2002}.
Our FWM spectra are well described by models incorporating these vibrational assignments, while additionally exposing their nonlinear couplings and coherence properties inaccessible to linear probes.
These modes furthermore reflect well-defined structural regimes governed by PbO content providing a structural framework for interpreting the FWM response:
from a SiO$_2$-dominated network below 35~mol\%, through a medium-range order (MRO) regime at 35--50~mol\%, characterized by growing Pb--Pb correlations and Pb-rich network regions, to a depolymerized network above 50~mol\%, where Pb--O--Pb free-oxide linkage become increasingly prevalent \cite{Takaishi2005, Alderman2021, Feller2010, Kacem2017} -- though we note that the precise compositional boundaries of these regimes and the nature of the Pb--O--Pb connectivity remain subjects of ongoing debate \cite{Alderman2021,Sen2024}.

These results establish THz-driven FWM as a powerful, selective technique for probing vibrational coherence, mode coupling, and dephasing dynamics in amorphous materials -- capabilities that remain inaccessible to linear spectroscopies.
Crucially, the technique is readily implemented using table-top laser systems, requiring no large-scale facility THz sources, which greatly lowers the barrier to adoption and positions THz-driven FWM as a broadly accessible tool for the ultrafast spectroscopy community.

\section{Results}

\subsection{Signal emergence and spectral characteristics}

Before presenting the time- and frequency-resolved FWM signals, we first establish the spatial origin of the measured response.
In our reflection geometry, the relevant wavevector mismatch for the FWM process is
\begin{equation}
    \Delta k(2\omega\pm\Omega)
    =
    k(2\omega\pm\Omega) + 2k(\omega) \pm k(\Omega)
\end{equation}
Since $\Omega\ll2\omega$, the $\Omega$-dependent term is suppressed relative to the leading term by a factor of $\Omega/2\omega\sim 10^{-3}$ (at 1~THz vs. the second-harmonic at 400~nm) regardless of $n(\Omega)$.
We therefore approximate $\Delta k\approx (2\omega/c)[n(\omega)+n(2\omega)]$, giving a coherence length
\begin{equation}
    L_{\rm coh}
    \approx
    \frac{\pi}{\Delta k}
    =
    \frac{\lambda}{4[n(\omega)+n(2\omega)]}
    \label{eq:interactionlength}
\end{equation}
Using manufacturer Sellmeier data for each glass composition (Supplementary Table~S1), this gives $L_{\rm coh}\approx51-60$~nm across our compositional series, confining the coherently-generated FWM signal to a near-surface layer of this thickness.
This interaction length is far shorter than the absorption lenghth of either the THz or near-IR fields in these materials (typically several $\mu$m or more), so the driving field amplitudes can be treated as effectively constant over $L_{\rm coh}$.

\begin{figure*}[ht]
\centering
\includegraphics[width=\linewidth]{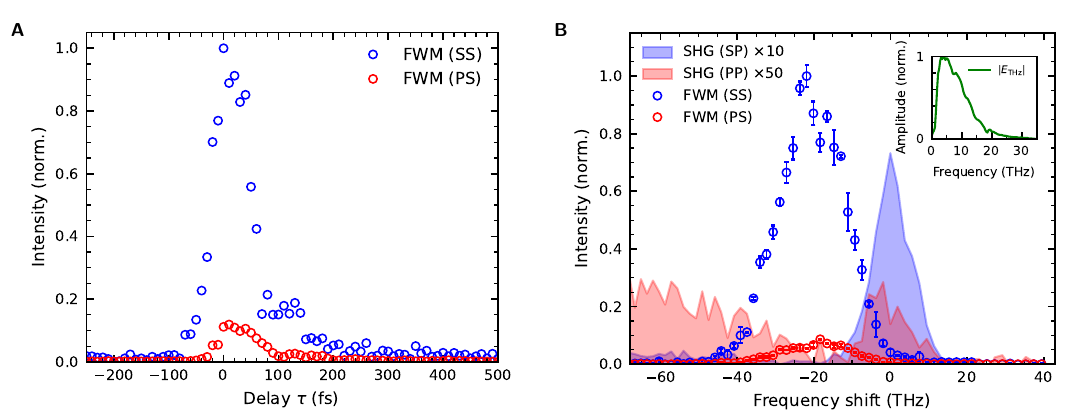}
\caption{
\textbf{FWM and SHG characterization of a lead glass sample with 44~mol\% PbO content.}
Only polarization configurations that exhibit nonzero responses are shown; combinations yielding no measurable signal are omitted for clarity.
The absence of signal in these configurations is consistent with the symmetry of the third-order nonlinear susceptibility tensor for isotropic, centrosymmetric media.
(\textbf{A}) Temporal evolution of the FWM signals at 400~nm reflected from the glass surface.
(\textbf{B}) FWM (circles; recorded at $\tau=0$) and SHG (shaded areas) spectra versus the frequency shift relative to the SHG carrier frequency (corresponding to 400~nm in wavelength). The error bars indicate the standard deviation from three consecutive measurements.
The inset displays the spectral amplitude of the THz pump pulse.
}
\label{fig:waveform-spectra-SF57}
\end{figure*}

Figure~\ref{fig:waveform-spectra-SF57}A shows the temporal evolution of the FWM signals at 400~nm, recorded in reflection from the top bulk layer of a lead glass sample containing 44~mol\% PbO for polarization configurations that yield measurable signals.
The first and last letters (S or P) of the polarization labels refer to the polarization states of the near-infrared (NIR) beam and the detected nonlinear signal, respectively. The THz polarization was fixed to S and is omitted from the notation for brevity.
The time-resolved signals both exhibit an impulsive response centered at zero delay, accompanied by a weak, asymmetric shoulder at positive delay times of approximately 130 fs. The dominant temporal features exhibit a full width at half maximum of about 80 fs, in agreement with the cross-correlation of the excitation pulses and indicative of an electronic third-order nonlinear response that is effectively instantaneous on the timescale of the optical excitation.

Figure~\ref{fig:waveform-spectra-SF57}B presents the FWM spectra (circles) recorded at zero-delay ($\tau=0$), where the NIR and THz pulses temporally overlap, along with the corresponding surface SHG spectra (shaded area, scaled by a factor of 10 and 50, respectively).
The frequency shift on the x-axis is defined relative to the fundamental SHG carrier frequency (corresponding to 400~nm in wavelength), enabling direct comparison of spectral features around the expected central emission.

In the absence of the THz field, weak surface SHG spectra, centered around 0~THz, are observed for the SP and PP configurations.
Additional spectral intensity is observed below $-15$~THz, which is attributed to residual supercontinuum light generated in the sample that could not be fully suppressed through spatial and spectral filtering.

Upon introduction of the THz field, strong and broadband FWM signals appear in the SS and PS configurations.
Particularly, the SS FWM signal exceeds the strongest SHG signal by more than an order of magnitude, with an enhancement factor greater than 13.
For these measurements, active baseline subtraction of the gated boxcar signal was applied, effectively removing contributions from surface SHG and residual supercontinuum.

Notably, the broadband FWM signals exhibit slight asymmetry and are highly Stokes-shifted, appearing at $2\omega - \Omega$, while no corresponding anti-Stokes signals at $2\omega + \Omega$ are detected.
Relative to the surface SHG spectra, the FWM peaks are red-shifted by approximately $-20$~THz, corresponding to a wavelength shift from 400~nm to approximately 406~nm.
Given that the THz pulse spectrum is centered around 5~THz, as shown in the inset of Figure~\ref{fig:waveform-spectra-SF57}B, the observed redshift cannot be explained solely to the spectral content of the driving THz field.
Moreover, while the FWM spectra extend beyond $-40$~THz, the THz pulse contains negligible spectral intensity above 25~THz.
This indicates that the broad FWM response arises from nonlinear interactions -- such as vibrational mode coupling -- rather than direct spectral overlap of the driving fields.

Our model of the third-order nonlinear response, which is presented below, links the pronounced redshifts and spectral broadening directly to coupling to low-frequency vibrational modes in the glass, including both collective Boson modes and localized vibrations such as Pb--O stretching modes.
These modes enhance the nonlinear response through vibrational resonance or dispersive contributions to the third-order susceptibility $\chi^{(3)}(2\omega\pm\Omega)$.
Furthermore, the model explains the strongly enhanced Stokes and the absent anti-Stokes sidebands: In our experiment, electronic states are populated exclusively through two-photon absorption (TPA) of the NIR pulse. The TPA resonance of the studied lead glass samples is at roughly 2.3~eV (539~nm) \cite{Tanaka2004}.
Therefore, the NIR photon energy $\hbar\omega_o$ of 1.55~eV (800~nm) places the two-photon excitation at 3.1~eV, approximately 0.8~eV above the TPA resonance.
Consequently, the Stokes-shifted emission at $2\omega_o-\Omega$ is driven closer to the TPA resonance, resonantly enhancing the Stokes response, while the anti-Stokes emission at $2\omega_o+\Omega$ is pushed further off-resonance and thus reduced.
This asymmetry is quantitatively captured by the third-order response function derived in Section~\ref{sec:theory}.

\subsection{Symmetry considerations, polarization dependence, and nonlinear scaling}

The observed polarization dependence of the FWM response (Figure~\ref{fig:waveform-spectra-SF57}A) is consistent with the symmetry constraints of third-order nonlinear processes in amorphous media, such as glass, which are statistically isotropic and centrosymmetric media. 
The relevant third-order susceptibility tensor components in this case are $\chi^{(3)}_{xxxx}$, $\chi^{(3)}_{xxyy}$, and $\chi^{(3)}_{xyyx}$ \cite{Shen1984book,Boyd2020book}, where the $z$-axis is defined along the surface normal, and the $y$-axis belongs to the plane of incidence.
Accordingly, only the SS and PS polarization configurations -- corresponding to the $\chi^{(3)}_{xxxx}$ and $\chi^{(3)}_{xyyx}$ tensor elements, respectively -- are expected to produce measurable FWM signals. 
This is in excellent agreement with the experimental polarization dependence we observed.
The SHG polarization dependence is similarly consistent with the expected surface susceptibility tensor decomposition involving $\chi^{(2),\text{s}}_{zxx}$, $\chi^{(2),\text{s}}_{xxz} = \chi^{(2),\text{s}}_{xzx}$, and $\chi^{(2),\text{s}}_{zzz}$ \cite{Shen1984book}.

The measured SS-to-PS signal ratio $I_{\rm SS}/I_{\rm PS}$ is approximately 12, somewhat above the expected value $\lvert\chi^{(3)}_{xxxx} / \chi^{(3)}_{xyyx}\rvert^2 \approx 9$, which assumes the standard Kleinman-like relation $\chi^{(3)}_{xxxx} = 3\chi^{(3)}_{xxyy} = 3\chi^{(3)}_{xyxy} = 3\chi^{(3)}_{xyyx}$, valid strictly in the off-resonance limit.
However, as noted above, the TPA resonance of this lead glass lies near 2.3~eV (539~nm) \cite{Tanaka2004}, placing the present optical excitation conditions in proximity to a two-photon electronic resonance.
Under such near-resonant conditions, the individual $\chi^{(3)}$ tensor components acquire distinct resonance-enhancement factors, lifting the degeneracy assumed by Kleinman symmetry and rendering the ratio $\lvert\chi^{(3)}_{xxxx} / \chi^{(3)}_{xyyx}\rvert^2$ material- and frequency-dependent.

The resonant dispersion of $\chi^{(3)}$ tensor components near electronic resonances is well established, having first been demonstrated experimentally by Yablonovitch, Bloembergen, and Wynne in $n$-InSb \cite{Yablonovitch1971} and subsequently confirmed in molecular systems by Levenson and Bloembergen \cite{Levenson1974}, who measured $\chi^{(3)}_{1111} / \chi^{(3)}_{1221}\approx1.84/0.54$ in benzene at 545~nm (an isotropic medium sharing the same $\chi^{(3)}$ tensor symmetry as glass, and probed near electronic resonance) yielding an intensity ratio of approximately 11.6, in close agreement with our observation.

One might ask whether this deviation instead reflects bulk electronic anisotropy intrinsic to the glass network, or whether cascaded $\chi^{(2)}$ contributions -- which, as shown by Flytzanis and Bloembergen \cite{Flytzanis1976}, can couple off-diagonal $\chi^{(3)}$ components to $[\chi^{(2)}]^2$ -- play a role.
The latter can be ruled out here: The SHG signal measured from the same samples is negligibly small, confirming that cascaded second-order contributions to the FWM response are negligible and that the measured ratio reflects the intrinsic bulk $\chi^{(3)}$ tensor.
Similarly, the SHG SP-to-PP ratio is not a straightforward diagnostic for bulk electronic anisotropy, since the PP channel involves multiple $\chi^{(2),\text{s}}$ tensor elements subject to Fresnel-weighted interference \cite{Shen1984book}.
Whether the FWM ratio encodes structural anisotropy beyond the near-resonant dispersion effect of the $\chi^{(3)}$ tensor cannot therefore be resolved from a single composition, and this question is revisited below in light of the systematic compositional dependence of the ratio.

\begin{figure}[ht]
\centering
\includegraphics[width=\linewidth]{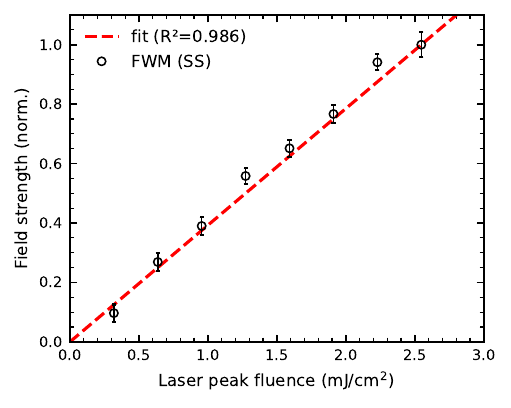}
\caption{
\textbf{FWM signal dependence on optical peak fluence in a lead glass sample with 44~mol\% PbO content.}
Each data point (circles) represents the mean field strength (square root of the intensity) across the full recorded spectrum at a given NIR laser peak fluence in the SS polarization configuration. The mean field strength is normalized to the maximum value, and error bars represent the weighted standard deviation across the spectrum. 
The red dashed curve shows a linear fit.
}
\label{fig:efficiency-SF57}
\end{figure}

We checked that the FWM process takes place in the perturbative regime by measuring its spectrum as a function of the NIR peak fluence by varying the pulse energy.
Figure~\ref{fig:efficiency-SF57} shows the mean intensity across the recorded spectrum (circles) for each fluence, measured in the SS polarization configuration from the lead glass sample containing 44~mol\% PbO.
Each value is normalized to the maximum mean intensity.
The error bars denote the weighted standard deviation across the spectrum, reflecting spectral variation.
The FWM response in field amplitude is seen to scale linearly with the driving NIR fluence $\mathcal{F}\propto\abs{E_{\rm las}}^2$ without notable saturation or distortions, confirmed by the excellent linear fit (red dashed curve) of the form $\abs{E_{\rm FWM}(\mathcal{F})} \propto \mathcal{F}$ resulting in a coefficient of determination $R^2=0.986$.
Further, this is consistent with a third-order nonlinear process governed by $\chi^{(3)}(2\omega\pm\Omega)$.
This confirms that the measured polarization ratios reflect the intrinsic $\chi^{(3)}$ response in the perturbative regime, uncontaminated by saturation or higher-order field effects.

Taken together, the polarization selection rules, tensor symmetry, and perturbative scaling confirm that the THz-driven nonlinear response originates from a bulk third-order FWM process, and establish the SS-to-PS ratio as a probe of the $\chi^{(3)}$ tensor, whose compositional evolution is examined in the following section.

\subsection{FWM spectra as a function of PbO content}\label{sec:FWM_spectra}

\begin{figure*}[ht]
\centering
\includegraphics[width=\linewidth]{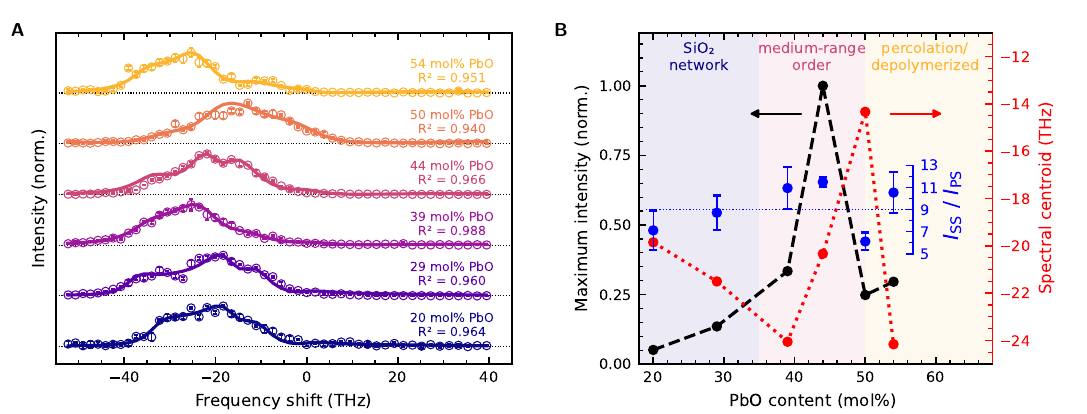}
\caption{
\textbf{PbO-dependent FWM response in lead glass samples.}
(\textbf{A}) FWM spectra in the SS polarization configuration of samples with varying PbO content plotted as a function of frequency shift relative to the SHG carrier frequency. The circles show the experimental data, with error bars indicating the standard deviation from three consecutive measurements. Solid lines represent modeled spectra (see Section 2.5) fitted to the data. All spectra are normalized and vertically offset for clarity.
(\textbf{B}) Maximum FWM signal intensity (black dashed, left axis), spectral centroid (red dotted, right axis), and $I_{\rm SS}/I_{\rm PS}$ polarization ratio (blue dots, inset axis) as a function of PbO content. The maximum FWM intensity is normalized with respect to the maximum signal observed at 44~mol\% PbO. The spectral centroid was calculated as the intensity-weighted average frequency shift relative to the SHG carrier frequency.
The $I_{\rm SS}/I_{\rm PS}$ ratio is defined as the ratio of the maximum FWM intensity measured in the SS configuration to that in the PS configuration.
Error bars indicate the standard deviation from three consecutive measurements.
Shaded regions indicate structural regimes as a function of PbO content: SiO$_2$ network ($<$35~mol\%), Pb--O--Pb chains (35–50~mol\%), and depolymerized ($>$50~mol\%).
}
\label{fig:spectra-centroid-PbO-content}
\end{figure*}

To investigate how the nonlinear response evolves with composition, we measured the FWM spectra for a series of lead glasses with increasing PbO content.
Figure~\ref{fig:spectra-centroid-PbO-content}A presents the normalized SS spectra (circles) recorded at fixed excitation conditions ($\sim$2.5~mJ/cm$^2$) for six compositions ranging from 20~mol\% to 54~mol\% PbO.
Each spectrum is plotted as a function of the frequency shift relative to the SHG carrier frequency.
A strong signal emerges in all cases on the Stokes side (at negative shifts).
While the overall spectral shape and amplitude vary considerably with PbO concentration, each spectrum extends beyond the spectral components of the THz pulse, indicating nonlinear coupling to vibrational modes.

To quantify the evolution of the FWM response with increasing PbO content, we extracted the maximum intensity and spectral centroid (i.e., the intensity-weighted average frequency shift) of the measured spectra, shown in Figure~\ref{fig:spectra-centroid-PbO-content}B for the SS polarization configuration.
The FWM intensity increases progressively between 20 and 40~mol\% PbO, followed by a sharp rise to a pronounced maximum at 44~mol\% PbO, before declining abruptly and stabilizing beyond 50~mol\%.
Meanwhile, the spectral centroid exhibits a non-monotonic trend.
Initially, the FWM spectrum redshifts with increasing PbO and reaches a minimum near 40~mol\% PbO. Beyond this point, the spectrum blueshifts toward 50~mol\% PbO, before a final sharp redshift at 54~mol\% PbO.
The PS configuration spectra (not shown) follow a similar trend in both maximum intensity and spectral centroid, although the intensity is roughly an order of magnitude lower.
These trends serve as a spectroscopic fingerprint of the underlying structural evolution within the glass network. 
To understand this, we define three structural regimes based on PbO content, informed by previous studies:
(i) a SiO$_2$ network regime ($<$ 35~mol\%), where Si--O--Si linkages dominate and Pb disrupts the network without notable spatial correlation;
(ii) a MRO regime (35--50~mol\%) characterized by emerging Pb--Pb correlations, edge-sharing PbO$_n$ units, and mixed Pb--O--Si / Pb--O--Pb connectivity; and
(iii) a percolation regime ($>$ 50~mol\%), marked by increasing Pb--O--Pb connectivity and depolymerization of the Si--O network, leading to structural percolation.
This classification is broadly consistent with structural probes (diffraction and EXAFS), spectroscopic techniques (NMR, Raman, and IR), and modeling results \cite{Takaishi2005,Rybicki2001,Alderman2021,Sen2024,Lee2014,Kacem2017,Drewitt2022,Feller2010}, though the precise boundaries should be treated as approximate.
In particular, diffraction-based models suggest the plumbite subnetwork may already be close to percolating at 35~mol\% PbO \cite{Alderman2021}, while $^{207}$Pb NMR measurements indicate that significant Pb--O--Pb free-oxide linkages, characteristic of the percolation regime, do not appear until above approximately 60~mol\% PbO \cite{Sen2024}.
This provides a structural framework for interpreting the recorded FWM spectra.
Particularly, the enhanced FWM signal observed at 44~mol\% PbO coincides with the region of evolving MRO in the glass network.
At this composition, neutron and X-ray diffraction studies indicate the emergence of well-defined Pb--Pb correlations around 3.8--4.0~\AA{}, consistent with the growth of a Pb-rich network \cite{Takaishi2005,Alderman2021,Kohara2010}.
MD simulations and EXAFS results further confirm that the corner-sharing connectivity of the SiO$_4$ tetrahedra network breaks near this composition \cite{Rybicki2001}, suggesting enhanced structural coherence and vibrational localization.
The nature of the Pb--O--Pb connectivity in this regime, however, is a matter of ongoing debate.
While diffraction-based models have been interpreted as evidence of edge-sharing [PbO$_n$] units forming a continuous Pb network even at relatively low PbO contents \cite{Takaishi2005,Kohara2010}, recent $^{17}$O NMR spectra at 60-71~mol\% PbO directly resolve Pb--O--Pb linkages only in that higher composition range \cite{Lee2014}, and $^{207}$Pb NMR finds negligible change in the isotropic shift from 30 to 60~mol\% PbO, suggesting that the fraction of Pb--O--Pb free-oxide linkages remain small below approximately 60~mol\% \cite{Sen2024}.
Accordingly, near 44~mol\% PbO, Pb--O--Si linkages are expected to strongly dominate, with only an incipient onset of Pb--O--Pb connectivity \cite{Sen2024,Lee2014}.
This is further supported by a multi-spectroscopic study -- including $^{29}$Si NMR, IR, Raman, and time-of-flight mass spectroscopy -- which demonstrated that the growth of Pb--O--Pb motifs and increased depolymerization are reflected in the progressive intensification of low-frequency vibrational features (e.g., the $\sim$140~cm$^{-1}$ Raman band, attributed to symmetric Pb--O stretching in covalent lead-oxygen pyramidal structures) and the redshift of silicate bands \cite{Feller2010}.
The coexistence of Pb--O--Si and incipient Pb--O--Pb units near 44~mol\% is plausible causing an optimal MRO leading to the observed enhancement of the FWM signal at this composition.
The formation of such MRO motifs not only increases local asymmetry and vibrational coupling but may also support quasi-localized or mixed vibrational modes that resonantly increase the third-order nonlinear susceptibility.
In contrast, the reduced signal at lower and higher PbO contents likely reflects either insufficient connectivity (at low PbO content) or  depolymerization (at high PbO content), resulting in decreased mode coherence or fewer contributing pathways.

The structural evolution across these regimes is also reflected in the vibrational spectra of lead silicate glasses, as established by IR and Raman spectroscopy \cite{Feller2010,Zagrai2024,Kacem2017}.
At low frequencies ($<$8~THz), the Boson peak of silica near 2~THz progressively gives way to Pb--O stretching and Pb$^{2+}$ rattling modes as PbO content increases; these contributions overlap spectrally around 3~THz and cannot be cleanly separated \cite{Feller2010,Kacem2017}.
In the 8--22~THz region, delocalized Si--O--Si network and ring vibrations coexist with emerging Pb--O bending and stretching contributions, whose relative weight grows with PbO content \cite{Zagrai2024,Kacem2017,Feller2010}.
Within this region, IR spectroscopy additionally resolves Pb--O--Pb bending (12--16.5~THz) and partly covalent Pb--O stretching in [PbO$_4$] units (16.5--21.6~THz), whose intensity grows progressively with PbO content \cite{Zagrai2024}, alongside a broad IR envelope near 15~THz attributed to the rocking motion of Si--O--Si bridges connecting Q$^n$ units, where Q$^n$ denotes a silicate tetrahedral unit with $n$ bridging oxygens ($n=$ 0--4) \cite{Feller2010}.
Above 22~THz, the Q$^4$ Si--O--Si bending mode shifts toward the Pb--O band above 47~mol\% PbO \cite{Zagrai2024}, while the asymmetric Si--O stretching envelope progressively redshifts and weakens by 50~mol\% \cite{Kacem2017}, reflecting systematic network depolymerization as Q$^4$ units are replaced by lower-connectivity Q$^n$ species \cite{Feller2010}.
These assignments are summarized in Table~\ref{tab:mode-assigment} and provide the structural reference framework against which the vibrational modes extracted from the FWM fits are interpreted in the following subsection.
\begin{table*}
\centering
\footnotesize
\caption{
\textbf{Assignments of vibrational modes in PbO-silicate glasses.}
Wavenumber and frequency ranges are correlated with characteristic structural units, with references to prior IR \cite{Feller2010,Zagrai2024}, Raman \cite{Feller2010,Kacem2017}, and THz time-domain \cite{Wada2024} spectroscopic analyses.
The boundary at 3~THz between the Boson peak and Pb$^{2+}$ rattling is approximate; in lead silicate glasses these contributions overlap spectrally and cannot be cleanly separated \cite{Feller2010,Kacem2017}.
The double rule separates spectrally resolved mode assignments (above) from the composite high-frequency stretch band (below), which overlaps partially with the preceding row and encompasses multiple Q$^n$ contributions whose relative weights vary with PbO content.
}
\label{tab:mode-assigment}
\begin{tabular}{cccc}
\hline
Wavenumber (cm$^{-1}$) & Frequency (THz) & IR assignment    & Raman assignment \\
\hline
10--100   & 0.3--3      & \multicolumn{2}{c}{Boson peak} \\
100--250  & 3--7.5      & Pb$^{2+}$ rattling; Pb--O stretch in [PbO$_3$]  & Pb--O stretch in [PbO$_4$] \\
250--650  & 7.5--18     & \multicolumn{2}{c}{Si--O--Si network and ring vibrations} \\ 
400--550  & 12--16.5    & Pb--O--Pb/O--Pb--O bending  & --- \\
550--720  & 16.5--21.6  & Pb--O stretch in [PbO$_4$], partly covalent & --- \\
720--850  & 21.6--25.5  & Q$^4$ Si--O--Si bending in pointed oxygen  & Asymmetric Si--O stretching \\
\hline
\hline
750--1200 & 22.5--36      & \multicolumn{2}{c}{Si--O stretch in Q$^n$ units; Pb--O stretch in [PbO$_n$]; Si--O--Pb stretch} \\ 
\hline
\end{tabular}
\end{table*}

Beyond the intensity and spectral centroid, the SS-to-PS signal ratio carries complementary information about the symmetry of the nonlinear response.
As discussed above, the ratio $I_{\rm SS}/I_{\rm PS}\sim\lvert\chi^{(3)}_{xxxx} / \chi^{(3)}_{xyyx}\rvert^2$ is sensitive to the relative weight of diagonal and off-diagonal $\chi^{(3)}$ tensor components, and cascaded $\chi^{(2)}$ contributions have been ruled out.
In PbO-containing glasses, the nonlinear response is dominated by the highly polarizable Pb$^{2+}$ lone-pair electrons\cite{Dimitrov1993,Dimitrov1996I,Dimitrov1996II}, whose hyperpolarizability contribution to $\chi^{(3)}$ in oxide materials is strongly sensitive to the identity and coordination of the surrounding cations \cite{Adair1989}.
The spatial organization of these lone pairs evolves with composition as the structural evidence above suggests.
If this evolution modulates the anisotropy of the electronic charge distribution, the SS-to-PS ratio should reflect it independently of the overall signal intensity. The compositional dependence of this ratio is examined in the following section.

\subsection{Compositional dependence of the SS-to-PS ratio}

Figure~\ref{fig:spectra-centroid-PbO-content}B (blue curve) shows the SS-to-PS signal ratio $I_{\rm SS}/I_{\rm PS}$ as a function of PbO content.
The ratio exhibits a non-monotonic evolution that closely tracks the structural regimes identified above, supporting its interpretation as a probe of bulk electronic anisotropy tied to the organization of Pb$^{2+}$ lone pairs within the glass network.
Since cascaded $\chi^{(2)}$ contributions have been ruled out, the ratio directly reflects the intrinsic $\chi^{(3)}$ tensor anisotropy -- specifically the symmetry of the Pb$^{2+}$ lone-pair charge distribution rather than the overall magnitude of the susceptibility. 

In the SiO$_2$-network regime (20--35~mol\% PbO), the ratio increases smoothly with PbO content.
At these compositions, Pb$^{2+}$ ions occupy isolated sites within the silicate network, with their lone pairs residing in the natural voids of the host silicate structure rather than aggregating into correlated plumbite arrangements \cite{Alderman2021}.
Let us note that this picture is in agreement with both a network-forming interpretation \cite{Alderman2021} and the modifier-like depolymerization behavior observed by vibrational spectroscopy at low PbO contents \cite{Kacem2017,Feller2010}.
The gradual increase in the ratio reflects the progressive incorporation of polarizable Pb$^{2+}$ units \cite{Dimitrov1996I,Dimitrov1996II,Adair1989}, whose stereochemically active lone pairs incrementally break the electronic isotropy of the host network and enhance the relative weight of diagonal $\chi^{(3)}$ components as PbO content increases.

The ratio reaches a maximum near 44~mol\% PbO, coinciding with the onset of medium-range order identified in the intensity data. 
Around this composition, the SiO$_4$ corner-sharing network loses three-dimensional connectivity \cite{Rybicki2001} while the plumbite subnetwork (already nearly percolating at lower PbO contents \cite{Alderman2021}) develops sufficient Pb-rich network connectivity that the stereochemically active lone pairs begin to aggregate within plumbite voids rather than merely occupying isolated silicate-network voids \cite{Alderman2021,Alderman2013}.
Because the electron density of Pb$^{2+}$ arises from the mixing of Pb 6$p$ states into the filled antibonding combination of O 2$p$ and Pb 6$s$ states \cite{Watson1999}, this collective lone-pair organization is spatially directional rather than isotropic, and the formation of chain- and layer-like plumbite motifs at this composition preferentially confines the lone-pair electron density within quasi-planar arrangements.
This anisotropic lone-pair distribution enhances the in-plane nonlinear susceptibility component $\chi^{(3)}_{xxxx}$ relative to the out-of-plane mixing component $\chi^{(3)}_{xyyx}$, selectively amplifying the SS-to-PS ratio.
Since the ratio is sensitive to tensor symmetry,
this peak reflects a genuine change in the directionality of the electronic response rather than simply an increase in the number of polarizable units.

Beyond 44~mol\% PbO, the ratio drops sharply at 50~mol\%, where PbO becomes the dominant network former.
As Si--O constraints weaken and the silicate network depolymerizes, the spatial confinement of lone pairs is lost and the electronic response becomes more isotropic in three dimensions, reducing the anisotropy of the $\chi^{(3)}$ tensor and decreasing the SS-to-PS ratio accordingly.

The ratio partially recovers at 54~mol\% PbO, suggesting a reorganization of the Pb-rich network into new structural motifs.
At these compositions, PbO$_3$ pyramidal units and Pb-rich network regions become increasingly prominent \cite{Takaishi2005,Drewitt2022,Kohara2010}, although the recent $^{207}$Pb NMR study suggests that edge-sharing between PbO$_n$ pyramids characteristic of the PbO polymorphs is likely not significant until substantially higher PbO contents \cite{Sen2024}.
The directional constraints imposed by Pb$^{2+}$ lone pairs, together with the emergence of correlated void structures, may promote locally anisotropic electronic environments.
This recovery indicates that the ratio is sensitive not only to the amount of PbO but to the specific geometry of Pb coordination and lone-pair spatial organization.

Taken together, the compositional evolution of the SS-to-PS ratio mirrors the structural crossovers identified through diffraction, NMR, and spectroscopic studies, and confirms that the FWM polarization ratio provides a sensitive and non-invasive probe of bulk electronic anisotropy in lead silicate glasses.

\subsection{Spectral modeling and mode assignments}\label{sec:results-fitting}

The FWM spectra across all six compositions were modeled using the third-order perturbative response framework derived in Section~\ref{sec:theory}, in which the measured intensity corresponds to the modulus-squared coherent sum over all contributing Liouville pathways at zero-delay ($\tau=0$) evaluated at the anti-Stokes- and Stokes-shifted emission frequencies $2\omega_o \pm \Omega'$.
For computational convenience the model represents the broad continua of vibrational excitations in the amorphous network by discrete effective moves, which capture the dominant nonlinear pathways without implying strictly discrete eigenstates (i.e. they have non-zero linewidths).

Fits to the SS spectra (solid lines in Figure~\ref{fig:spectra-centroid-PbO-content}A) yield coefficients of determination $R^2$ between 0.94 and 0.99 across the full composition range.
The model accurately reproduces the strongly Stokes-shifted character of the spectra and the negligible anti-Stokes signal, which can be rationalized by interference among different $\chi^{(3)}$ pathways whose relative phases suppress anti-Stokes emission under the present excitation conditions \cite{McDonnell2024}.
\begin{figure*}[ht]
\centering
\includegraphics[width=\linewidth]{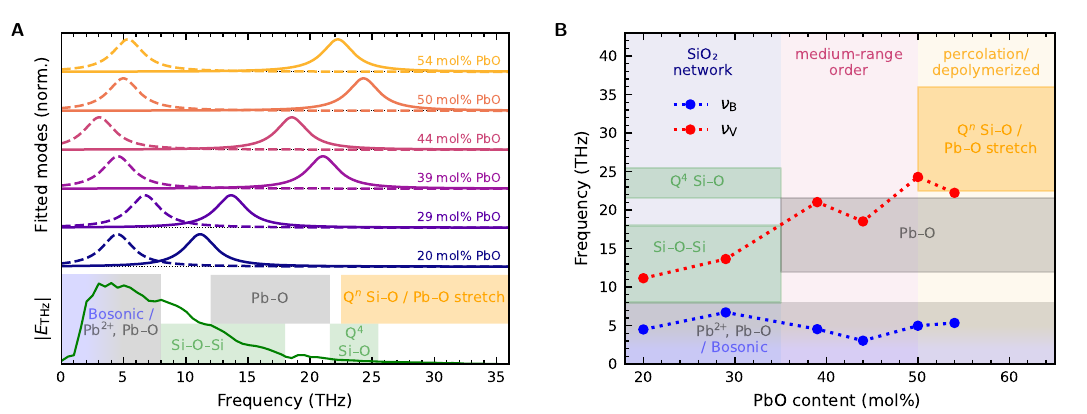}
\caption{
\textbf{Fitted vibrational response functions and extracted mode frequencies across PbO compositions.}
\textbf{(A)}
Imaginary parts of the bosonic (dashed) and vibrational (solid) Lorentzian response functions extracted from fits to the SS FWM spectra, plotted as a function of frequency for each composition (color scale, 20--54~mol\% PbO). Each mode is normalized to its peak amplitude. 
The green curve shows the spectral amplitude of the THz pump pulse; the extension of the extracted vibrational modes well beyond the spectral peak of the THz field reflects nonlinear coupling to vibrational frequencies outside the directly driven bandwidth.
Shaded regions indicate frequency ranges associated with known structural assignments in PbO-silicate glasses (see Table~\ref{tab:mode-assigment}).
\textbf{(B)}
Extracted bosonic ($\nu_{\rm B}$, blue) and vibrational ($\nu_{\rm V}$, red) peak frequencies as a function of PbO content, overlaid on the same structural band assignments as background shading. Shaded vertical regions indicate the three structural regimes: SiO$_2$ network ($<$35~mol\%), MRO (35--50~mol\%), and percolation ($>$50~mol\%).
}
\label{fig:fitted_modes}
\end{figure*}
A systematic analysis established that the vibrational linewidth $\gamma_{\rm vib}$ is not independently resolvable under broadband femtosecond excitation, i.e. variations within the physically plausible range produced negligible changes in the fit quality, with the optimization redistributing amplitude between $\mu_{\rm V}/\mu_{\rm B}$ and $\mu_{\rm VB}$ instead (here,  B denotes for a collective bosonic mode, and V represents a higher-frequency network stretching mode; see Table~\ref{tab:fit-parameters} for more details).
This parameter compensation is a consequence of the convolution-limited spectral resolution.
Accordingly, $\gamma_{\rm vib}$ was fixed at a representative intermediate value of 1.5~THz and $\mu_{\rm VB}$ was bounded within a geometrically motivated interval (0.1--10) anchored to $\sqrt{\mu_{\rm V}/\mu_{\rm B}}$, yielding a constrained parameterization in which the remaining free parameters track physically interpretable trends across composition.
Because the driving THz field carries negligible spectral weight above 30~THz, the FWM response is restricted to the sub-30~THz region, and since THz-driven FWM probes $\chi^{(3)}$, modes that are weak or silent in IR and Raman spectra may contribute to the response through nonlinear coupling \cite{McDonnell2024}.

The extracted bosonic and vibrational response functions are shown in Figure~\ref{fig:fitted_modes}A alongside the spectral regions associated with known structural assignments summarized in Table~\ref{tab:mode-assigment}.
All the fitted parameters are summarized in Table~\ref{tab:fit-parameters}.
\begin{table*}
\centering
\caption{
\textbf{Fitted parameters of the third-order nonlinear response
model across PbO compositions.}
The TPA detuning $\delta$ and electronic dephasing $\gamma_{\rm el}$ characterize the off-resonant two-photon electronic response.
The bosonic and vibrational mode frequencies $\nu_{\rm B}$ and $\nu_{\rm V}$ correspond to the peak frequencies of the fitted Lorentzian response functions associated with the collective low-frequency and network stretching contributions, respectively.
The ratio $\mu_{\rm V}/\mu_{\rm B}$ quantifies the relative coupling strength of the vibrational to the bosonic channel, and $\mu_{\rm VB}$ is the cross-manifold coupling amplitude, constrained to the interval $[0.1, 10]$ (see text).
Horizontal rules separate the three structural regimes: SiO$_2$ network ($<$35~mol\%), MRO (35--50~mol\%), and percolation ($>$50~mol\%).
Values are rounded to three significant figures.
}
\label{tab:fit-parameters}
\small
\begin{tabular}{@{}ccccccccc@{}}
\\
\hline
PbO (mol\%) & $R^2$ & $\delta$ (THz) & $\gamma_{\rm el}$ (THz) & $\nu_{\rm B}$ (THz) & $\nu_{\rm V}$ (THz) & $\mu_{\rm V}/\mu_{\rm B}$ & $\mu_{\rm VB}$ \\
\hline
20  &  0.964   & -21.6  & 4.54   & 4.50   & 11.2   & 0.49   & 10.0  \\
29  & 0.960    & -24.2  & 4.15   & 6.72   & 13.6   & 0.99   & 10.0  \\
\hline
39  & 0.988    & -19.0  & 6.33   & 4.55   & 21.0   & 1.05   & 0.33  \\
44  & 0.966    & -19.1  & 4.51   & 3.05   & 18.5   & 0.78   & 0.57  \\
\hline
50  & 0.940    & -26.0  & 7.97   & 4.99   & 24.3   & 0.38   & 0.90  \\
54  & 0.951    & -19.0  & 5.53   & 5.35   & 22.2   & 0.94   & 0.33  \\
\hline
\end{tabular}
\end{table*}
The bosonic mode (dashed lines) remains within the low-frequency band ($<$7.5~THz) encompassing the Boson peak and Pb$^{2+}$ rattling / Pb--O contributions across all compositions, consistent with a collective excitation that acquires increasing Pb--O character with PbO content.
The vibrational mode (solid lines), by contrast, undergoes a systematic blueshift with increasing PbO, made explicit in Figure~\ref{fig:fitted_modes}B, where the extracted frequencies $\nu_{\rm B}$ and $\nu_{\rm V}$ are plotted against composition with the same structural band assignments as background shading.
While $\nu_{\rm B}$ shows no discernible compositional trend, $\nu_{\rm V}$ rises monotonically from the Si--O--Si network band (7.5--18~THz) at low PbO concentrations into the composite Q$^n$ Si--O / Pb--O stretch band ($>$21.6~THz) at high PbO concentrations, crossing assignment boundaries near 35~mol\% and again near 50~mol\% PbO.
These crossings coincide with the structural regime boundaries identified above, directly linking the evolution of the dominant nonlinear response mode to the progressive transformation from a silicate-dominated network, where Pb$^{2+}$ acts as a modifier, toward a Pb-right network in which Pb--O and Si--O--Pb stretching modes increasingly govern the third-order susceptibility.
The electronic dephasing rate $\gamma_{\rm el}$ is elevated near 50~mol\% PbO, suggesting increased inhomogeneous broadening of the TPA resonance, consistent with the structural heterogeneity expected near the percolation threshold where Pb--O--Pb and Si--O--Si environments coexists.

\begin{figure}[ht]
\centering
\includegraphics[width=\linewidth]{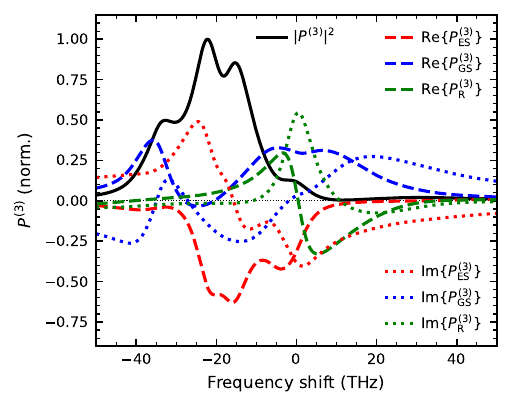}
\caption{
\textbf{Pathway-resolved decomposition of the fitted third-order polarization $P^{(3)}(\Omega')$ for the 44~mol\% PbO glass.}
The total fitted response $\abs{P^{(3)}(\Omega')}^2$ (black solid line) is decomposed into its excited-state (ES, red), ground-state (GS, blue), and THz-resonant (R, green) pathway contributions (Eqs.~\eqref{eq:amplitude-ES}-\eqref{eq:amplitude-R}), shown as real (dashed) and imaginary (dotted) parts.
All curves are normalized to the peak of $\abs{P^{(3)}(\Omega')}^2$.
This composition is shown as representative of the full series; equivalent decompositions for the remaining five compositions are provided in Supplementary Figure~S1.
The relative phase and magnitude of the three pathway classes account for the pronounced Stokes-shifted FWM signal and the suppressed anti-Stokes signal discussed in Section~\ref{sec:FWM_spectra}.
}
\label{fig:P3_decomposition}
\end{figure}
To illustrate how pathway interference shapes the lineshape, Figure~\ref{fig:P3_decomposition} decomposes the fitted $P^{(3)}(\Omega')$ for the 44~mol\% PbO sample into its ES, GS, and R contributions (Eqs.~\eqref{eq:amplitude-ES}-\eqref{eq:amplitude-R}).
The excited-state (ES) term dominates and is strongly negative on the Stokes side, while the ground-state (GS) and THz-resonant (R) terms grow in near zero THz and partially cancel the excited-state contribution on the anti-Stokes side, producing the pronounced Stokes/anti-Stokes asymmetry seen in $\abs{P^{(3)}(\Omega')}^2$.
Equivalent decompositions for all six compositions are provided in Supplementary Figure~S1, where the same qualitative interference pattern persists but the relative pathway weights evolve systematically with PbO content.

\section{Discussion}

In conclusion, we have demonstrated that THz-driven FWM at glass surfaces provides a direct and sensitive probe of low-frequency vibrational modes and structural evolution in disordered materials.
Applied to a compositional series of PbO--silicate glasses, this approach reveals distinct spectral signatures associated with collective Boson-peak excitations, Pb--O stretching and bending modes, and Si--O--Si network vibrations, highlighting its ability to capture subtle variations in vibrational coherence and medium-range structural order.
Particularly, we can track their systematic evolution across three structurally distinct regimes (i.e. a polymerized SiO$_2$ network, medium-range order, and percolation / a depolymerized SiO$_2$ network) as a function of PbO content.
The monotonic blueshift of the dominant vibrational frequency $\nu_{\rm V}$ from the Si--O--Si network band into the composite Pb--O / Q$^n$ stretch band directly reflects the progressive transformation of the Pb$^{2+}$ network role -- from silicate-network--dominated environments at low PbO content, where Pb primarily depolymerizes the Si--O backbone, toward a Pb-rich network at high PbO content, where Pb--O and Si--O--Pb stretching modes increasingly govern the nonlinear response,
while the pronounced enhancement of the FWM signal near 44~mol\% PbO is linked to the emergence of Pb--Pb correlations and a structural transition marked by growing Pb-rich network regions alongside dominant Pb--O--Si connectivity.
The nonlinear polarization ratio $I_{\rm SS}/I_{\rm PS}$ provides a complementary probe of $\chi^{(3)}$ tensor anisotropy, whose compositional evolution mirrors the reorganization of Pb$^{2+}$ lone-pair spatial correlations across the same structural regimes.
Together, these observables establish THz-driven FWM as a depth-confined probe of the near-surface region (probe depth $\sim$50~nm, set by the reflection-geometry wavevector mismatch, Eq. \eqref{eq:interactionlength}), vibrationally sensitive to structural correlations characteristic of the underlying glass network.
The compositional trends we extract -- the systematic blueshift of vibrational frequencies with PbO content, the non-monotonic anomaly at 44~mol\% PbO -- track closely with bulk-averaged IR, Raman, and THz-TDS assignments from prior work \cite{Feller2010,Zagrai2024,Kacem2017,Wada2024}, supporting a bulk-representative interpretation of the probed layer while further resolving vibrational and electronic structural fingerprints that cannot be individually resolved in linear IR, Raman, and THz-TDS measurements.

Several limitations of the present study should be acknowledged.
The broadband femtosecond excitation employed here imposes a fundamental convolution limit on the spectral resolution, rendering individual modes and dephasing rates (particularly $\gamma_{\rm v}$) non-identifiable from the data and requiring their fixation at physically motivated values.
The response model, while capturing the essential nonlinear pathways, represents the broad continua of vibrational excitations in the amorphous network by discrete effective modes, which limits the ability to resolve overlapping contributions within the same spectral band, most notably the Boson peak and Pb$^{2+}$ rattling modes below 7.5~THz.
The FWM signal originates from approximately the top 50~nm of the bulk, depending on composition (Eq. \eqref{eq:interactionlength}).
Samples where used as received with their standard manufacturer polish and no additional surface preparation.
A possible concern is surface de-alkalization or lead leaching, a known aging process in lead-silicate glasses that preferentially removes network-modifying Pb and can produce a near-surface layer with reduced refractive index and Pb content \cite{Schultz-Munzenberg1998}.
Notably, in that study SF57 (44~mol\%) -- the composition showing our strongest and most structurally distinctive FWM response -- was identified as the most leaching-prone of the lead silicates examined, owing to its low alkali content.
Substantial Pb depletion within our probed depth would therefore be expected to suppress, not enhance, the Pb-related spectral signatures we observe; instead, we find the opposite, with the clearest lone-pair-driven $\chi^{(3)}$ anisotropy occurring precisely at this composition and coinciding with an independently established bulk structural transition.
This is consistent with our signal reflecting genuinely bulk-like Pb coordination rather than a depleted surface layer.
We note further that even the residual network-forming Pb reported to survive leaching \cite{Schultz-Munzenberg1998} resides within a restructured, densified silicate network (i.e. leaching drives a marked shift in Si Q$^n$-speciation) rather than the intact medium-range scaffold we argue is necessary to confine and order the lone-pair correlations underlying our signal; a leached layer is therefore not expected to reproduce this feature even where some network-forming Pb remains.
Finally, the current analysis is restricted to the SSS polarization configuration for the compositional study; a full polarization-resolved decomposition of the $\chi^{(3)}$ tensor across compositions would provide a more complete picture of the anisotropy evolution.

Beyond their role as a structural probe, the present results contribute new experimental evidence to the ongoing debate on the compositional evolution of the Pb$^{2+}$ network role in lead silicate glasses.
The debate has centered on when, and in what sense, Pb$^{2+}$ transitions from silicate-network--modifying to network-forming behavior, with diffraction-based models arguing for a network-forming role over the entire glass-forming range \cite{Alderman2021}, while NMR \cite{Sen2024} and vibrational spectroscopy \cite{Feller2010,Kacem2017} describe subtler, more gradual changes in local structure and connectivity.
The nonlinear optical observables reported here add a dimension that none of these techniques directly accesses: the compositional evolution of the bulk $\chi^{(3)}$ tensor and, through it, the spatial organization of Pb$^{2+}$ lone-pair electron density.
The non-monotonic peak in both FWM signal intensity and $I_{\rm SS}/I_{\rm PS}$ ratio at 44~mol\% PbO (i.e. a composition where NMR finds no dramatic change in local Pb--O coordination \cite{Sen2024} and where free-oxide Pb--O--Pb linkages are negligible \cite{Lee2014,Sen2024}) provides direct evidence that a collective reorganization of lone-pair spatial correlations occurs in the medium-range structure at this composition, independently of changes in nearest-neighbour bonding.
This supports the picture proposed by Alderman et al. in which the growing directionality and aggregation of lone pairs within the plumbite voids is a structurally distinct process from changes in Pb--O bond character \cite{Alderman2021,Alderman2013}.
The subsequent sharp drop in the ratio above 44~mol\% PbO, coinciding with Si--O network depolymerization and a more three-dimensionally isotropic Pb environment \cite{Kacem2017,Sen2024}, further suggests that lone-pair spatial order is disrupted by the loss of the silicate scaffold that confines the plumbite voids.
Notably, 44~mol\% PbO falls close to the compositions independently proposed as marking a related structural transition \cite{Wang1996,Meneses2006} (directly contested by Alderman et al. \cite{Alderman2021} on the grounds that local Pb-O coordinate shows no corresponding change in this region) and consistent with the silicate-network connectivity break reported independently from molecular dynamics/EXAFS \cite{Rybicki2001}.
Our results suggest these need not be in conflict: a genuine structural transition may indeed occur near this composition, but one expressed primarily through medium-range lone-pair correlation rather than local coordination geometry -- a degree of freedom invisible to diffraction-derived coordination numbers and NMR chemical shifts alike, but to which the present $\chi^{(3)}$ observables are directly sensitive.
These observations suggests that THz-driven FWM may offer a general route to disambiguating structural-role debates of this kind in other lone-pair or polarizable-cation glass systems.

More broadly, these findings suggests that THz-driven FWM could become a valuable tool for understanding and engineering amorphous materials, providing direct spectroscopic access to the low-frequency vibrational dynamics -- Boson-peak excitations, network stretching modes, and their coupling -- that govern mechanical damping, thermal conductivity, and nonlinear optical response in disordered solids.
The sensitivity of the FWM signal and polarization ratio to medium-range structural order and lone-pair organization suggests that the technique could be extended to other disordered materials, where analogous questions about the relationship between local structure, vibrational coherence, and macroscopic properties remain open.

Future work could pursue several directions.
Replacing the femtosecond optical probe with narrowband picosecond pulses would directly lift the convolution limit on the detection side, enabling independent determination of vibrational dephasing rates and potentially resolving overlapping contributions within the same spectral band -- most notably the Boson peak and Pb$^{2+}$ rattling modes below 7.5~THz.
Extension to other technologically relevant glass families (chalcogenides, heavy-metal oxides, or phase-change materials) would test the generality of the vibrational mode migration observed here as a structural probe.
Finally, combining THz-driven FWM with complementary structural probes such as neutron diffraction or solid-state NMR on the same sample series would enable direct correlation of the fitted vibrational parameters with independently determined structural descriptors, strengthening the physical interpretation of the extracted mode frequencies and their compositional evolution.
Beyond bulk-representative structural characterization, the technique's intrinsic near-surface confinement ($\sim$50~nm) could be deliberately leveraged to study surface-localized phenomena in their own right -- such as leaching, weathering, or polishing-induced alteration layers in technologically relevant glasses -- where depth-resolved sensitivity would be the primary asset.

\section{Materials and Methods}

\subsection{Experimental design}

The experiments were performed using a Ti:Sapphire regenerative amplifier operating at a 1-kHz repetition rate, producing 50-fs, linearly polarized, NIR optical pulses centered at 800~nm with an energy of 3~mJ per pulse.
The experimental setup resembled a common pump-probe configuration. A schematic can be found in our previous work \cite{Dalstein2024}.

A portion of the beam (1.3~mJ) was used to generate intense ultrashort THz pulses via two-color plasma generation in air, serving as the pump beam.
These THz pulses ($\sim$60~fs full width half maximum duration) were collimated and focused onto the sample at a 45$^\circ$ incidence angle using a pair of off-axis parabolic (OAP) mirrors. 
A 1-mm-thick high-resistivity float-zone silicon wafer filtered out residual optical beams from the THz path, while a wire grid THz polarizer ensured S-polarized THz pulses at the sample surface, with an estimated peak electric field of $\sim$340~kV/cm.

The probe beam, another small portion of the initial NIR beam (10~$\mu$J), was collinearly focused and spatially overlapped with the THz beam on the sample through a hole in the second OAP mirror using a 400-mm focal length lens.
A half-wave plate controlled the linear polarization state of the NIR beam, and a motorized translation stage ensured precise control over the temporal overlap between the NIR and THz pulses.

The reflected FWM signal, generated at the surface (top bulk layer) of the sample, was filtered by a colored glass filter to remove the residual NIR beam. 
It was then analyzed and recorded using a polarizer and a monochromator equipped with a photomultiplier tube at the output. 
A gated boxcar averager, combined with an optical chopper in the THz path operating at 500 Hz, allowed us to selectively record the THz-driven FWM signal, eliminating background contributions and improving the signal-to-noise ratio (active baseline subtraction).
Each recorded data point was the average of 150 pulses.

For the NIR excited surface SHG signals, each data point represented the average of 300 pulses as the active baseline subtraction was not implemented for this single-beam experiment.

\subsection{Samples}

The materials studied were samples of optics-grade lead oxide silicate glasses obtained from Schott (SF2, SF1, SF6, SF57, SF58, and SF59; PbO contents of 20, 29, 39, 44, 50, and 54~mol\%, respectively), used as received with the manufacturer's standard optical polish. No additional surface preparation was performed.

\subsection{Modeling of the third-order nonlinear polarization}\label{sec:theory}

To interpret the spectral features and Stokes/anti-Stokes asymmetry of the measured FWM response, we develop a third-order perturbative model of the nonlinear polarization driven by the combined NIR and THz fields.
The full derivation is provided in the Supplementary Material; here we summarize the physical picture and the key results used to fit the experimental spectra.

\begin{figure*}[!ht]
\centering
\includegraphics[width=\linewidth]{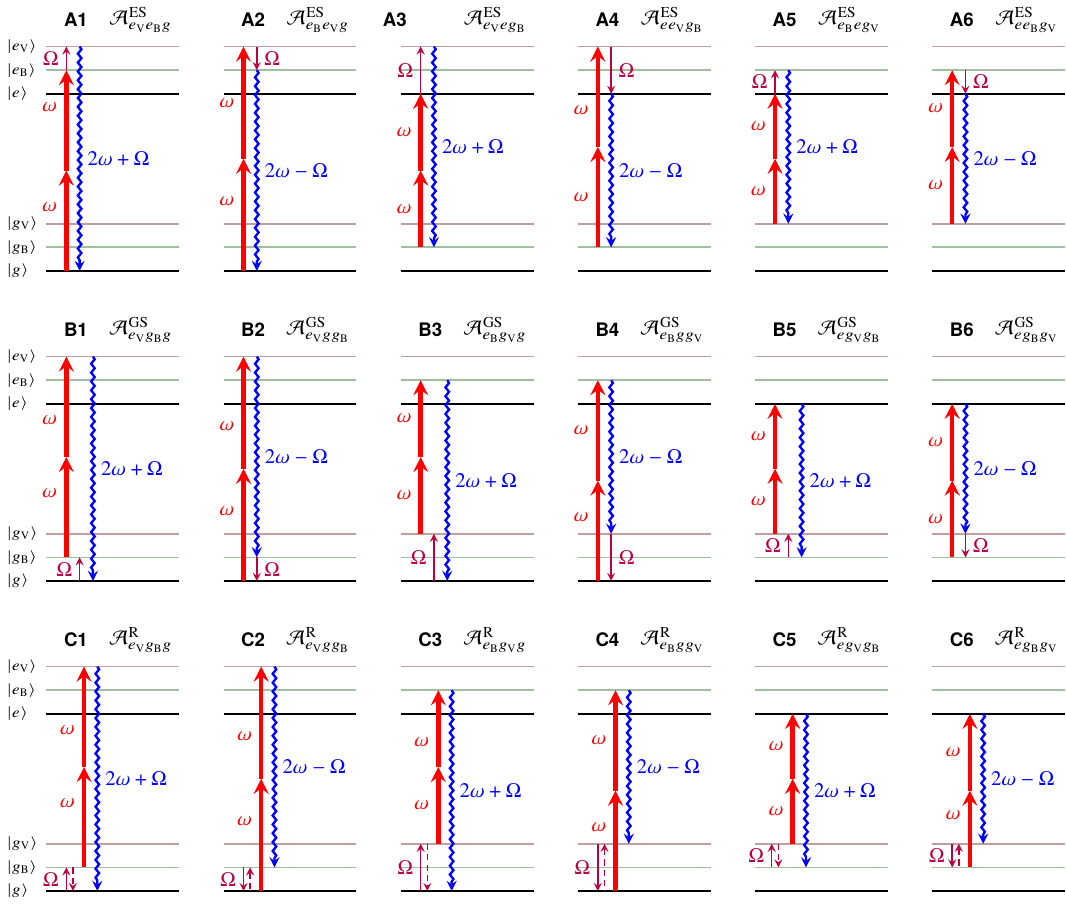}
\caption{
\textbf{Liouville pathways contributing to the THz-driven FWM response.}
The eighteen signal-active third-order pathways are grouped into excited-state (ES, \textbf{A1--A6}), ground-state (GS, \textbf{B1--B6}), and resonant (R, \textbf{C1--C6}) classes, 
distinguished by whether the THz interaction ($\Omega$, purple arrows) occurs on the ket-side within $\ket{e^\star}$ (ES) or $\ket{g^\star}$ (GS) after the optical TPA transition; or whether it drives a resonant ground-state intraband transition (R), which can occur on either the ket- or the bra-side.
}
The dashed arrows indicate that the THz field acts on the bra-side.
Red arrows: NIR interactions at $\omega$.
Blue wavy arrows: emitted FWM signal at $2\omega\pm\Omega$.
Within each group, odd- and even-numbered diagrams corresponds to anti-Stokes ($2\omega+\Omega$) and Stokes ($2\omega-\Omega$) emission, respectively.
\label{fig:pathways}
\end{figure*}

\subsubsection*{Level structure and interactions}

The glass is modeled as a system with a ground-state manifold $\ket{g^\star}=\{\ket{g}, \ket{g_{\rm B}}, \ket{g_{\rm V}}\}$ and an excited-state manifold $\ket{e^\star}=\{\ket{e}, \ket{e_{\rm B}}, \ket{e_{\rm V}}\}$, where the subscripts $b$ and $v$ denote states displaced by the Boson-peak ($\Omega_{\rm B}$) and network vibrational ($\Omega_{\rm V}$) mode frequencies, respectively (Figure~\ref{fig:pathways}).
The NIR field drives two-photon transitions between $\ket{g^\star}$ and $\ket{e^\star}$ at frequency $2\omega_o$, detuned from the TPA resonance by $\Delta<0$ (i.e. the NIR is above resonance).
The THz field drives transitions within $\ket{g^\star}$ and $\ket{e^\star}$, coupling the vibrational sublevels $\ket{g_{\rm B}}$ and $\ket{g_{\rm V}}$ to $\ket{g}$, and $\ket{e_{\rm B}}$ and $\ket{e_{\rm V}}$ to $\ket{e}$, respectively.

\subsubsection*{Contributing pathways}

Three classes of Liouville pathways contribute to the third-order polarization, distinguished by the interaction order of the THz field (Figure~\ref{fig:pathways}):
\textit{Excited-state (ES) pathways} with amplitude
\begin{equation}
    \mathcal{A}^{\rm ES}_{apb} = \mu_{ap}\mu^{(2)}_{pb}P_b
    \label{eq:amplitude-ES}
\end{equation}
where the system is promoted to $\ket{e^\star}$ by two NIR photons, then the THz field acts while the system resides in the excited state, and the signal is emitted upon return to $\ket{g^\star}$.
\textit{Ground-state (GS) pathways} with amplitude 
\begin{equation}
    \mathcal{A}^{\rm GS}_{aqb} = -\mu^{(2)}_{aq}\mu_{qb}P_q
    \label{eq:amplitude-GS}
\end{equation}
where the TPA transition again occurs first and the THz field acts via stimulated Raman-like scattering between ground-state sublevels; the vibrational resonance condition links the THz frequency to the optical detuning and emitted sideband.
\textit{THz-resonant (R) pathways} with amplitude 
\begin{equation}
    \mathcal{A}^{\rm R}_{aqb} = \mu^{(2)}_{aq}\mu_{qb}(P_b - P_q)
    \label{eq:amplitude-R}
\end{equation}
where the THz field resonantly drives a ground-state intraband coherence ( within $\ket{g^\star}$) before the two-photon optical interaction, allowing the coherence to build up under the resonance condition $\Omega\approx\omega_{qb}$; the amplitude is weighted by the population difference $(P_b - P_q)$, and the interaction may occur on either the ket- or bra-side.
In Eqs. \eqref{eq:amplitude-ES}--\eqref{eq:amplitude-R}, $\mu_{ap}$ is the one-photon transition dipole moment between states $a \in e^\star$ and $p \in e^\star \setminus\{a\}$, $\mu^{(2)}_{ab}$ is the effective two-photon transition dipole moment between $a \in e^\star$ and $b \in g^\star$, and $P_{b,q}$ are the thermal populations of the ground states $\ket{b}$ and $\ket{q \in g^\star\setminus\{b\}}$ following a Boltzmann distribution at room temperature.

\subsubsection*{FWM intensity and spectral response}

The measured FWM intensity corresponds to the modulus-squared coherent sum over all contributing pathways evaluated at the Stokes- and anti-Stokes-shifted emission frequencies $\omega' = 2\omega_o + \Omega'$:
\begin{align}
    I(\omega')
    =
    \abs{\tilde P^{(3)}(\omega')}^2
    =
    \abs{\sum_{ab} \tilde P^{(3)}_{ab}(2\omega_o + \Omega')}^2
    \label{eq:intensity}
\end{align}
where the sum runs over all signal-active coherence pathways $ab\in\{ e g_{\rm B},\, e g_{\rm V},\, e_{\rm B} g,\, e_{\rm B} g_{\rm V},\, e_{\rm V} g,\, e_{\rm V} g_{\rm B} \}$. Any SHG-like pathways (e.g. $eg$) are neglected.
The emitted sideband offset $\Omega'$ generalizes the nominal signal frequencies $2\omega_o \pm \Omega$, accounting for the fact that the measured spectrum reflects a coherent superposition over all contributing THz frequency components $\Omega$ within the pulse bandwidth.
Each pathway contribution takes the form
\begin{align}
    \tilde P^{(3)}_{ab}(2\omega_o + \Omega')
    =
    \mu_{ab}
    \int_0^\infty
    \frac{\dif\Omega}{2\pi}
    &\abs{\tilde E_\Omega(\Omega)}
    \mathcal{G}_{ab}(\Omega')
    \notag\\
    &\times
    \sum_{l=\pm 1}
    \mathcal{R}^{(l)}_{ab}(\Omega',\Omega)
    \label{eq:pathway}
\end{align}
Since all FWM spectra are recorded and fitted at zero pump--probe delay ($\tau=0$) with near--transform-limited THz pulses ($\phi_{\rm THz}(\Omega)\approx0$), the THz phase factor reduces to unity and the response functions depend only on the THz spectral amplitude $\abs{\tilde E_\Omega(\Omega)}$.
$\mathcal{R}^{(l)}_{ab}$ is a pathway-specific response function encoding the vibrational resonances (see Supplementary Material), and $\mathcal{G}_{ab}(\Omega')$ is the two-photon propagator
\begin{equation}
    \mathcal{G}_{ab}(\Omega')
    =
    \frac{(i/\hbar)^2}{\Gamma_{\rm el}+i(\Delta_{ab}-\Omega')}
    \label{eq:propagator}
\end{equation}
with $\Gamma_{\rm el}$ the electronic dephasing rate and $\Delta_{ab} = \Delta + \sigma\Omega_m$ the TPA detuning accounting for vibronic shifts ($\sigma \in \{1,-1\}$ and $\Omega_m \in \{\Omega_{\rm B}, \Omega_{\rm V}, \Omega_{\rm V}-\Omega_{\rm B}\}$).
Since $\Delta_{ab} < 0$ in the present experiment, the Stokes sideband ($\Omega' < 0$) is driven closer to the TPA resonance, enhancing $\abs{\mathcal{G}_{ab}}$, while the anti-Stokes sideband ($\Omega' > 0$) is pushed further off-resonance and consequently suppressed -- consistent with the observed spectral asymmetry.

The response functions $\mathcal{R}^{(l)}_{ab}$ encode two types of vibrational resonances:
optical sideband resonances through denominators of the form $[\Gamma_{\rm vib}+i(\Delta_{ab}-\Omega'+l\Omega - s\, \Omega_m)]^{-1}$ with $l,s \in \{1,-1\}$, which peak when the emitted frequency aligns with a vibronically shifted TPA resonance, and THz-driven vibrational resonances through denominators of the form $[\Gamma_{\rm vib} + i(s\,\Omega_m - l\Omega)]^{-1}$, which peak when a THz frequency component $\Omega$ matches a vibrational mode frequency $\Omega_m$.
Together these contribute to the pronounced redshift and broadening of the FWM spectra beyond the driving THz bandwidth.

\subsubsection*{Numerical implementation and free parameters}

For numerical fitting, the broad continua of vibrational excitations in the amorphous network are represented by two effective Lorentzian modes: a collective low-frequency bosonic mode at frequency $\nu_{\rm B}=\Omega_{\rm B}/2\pi$ capturing the Boson peak and Pb$^{2+}$ rattling / Pb--O contributions, and a higher-frequency vibrational mode at $\nu_{\rm V}=\Omega_{\rm V}/2\pi$ representing network stretching contributions.
The vibrational linewidth $\gamma_{\rm vib}=\Gamma_{\rm vib}/2\pi$ was fixed at 1.5~THz (see Section~\ref{sec:results-fitting} for justification).
The remaining free parameters are the TPA detuning $\delta = \Delta/2\pi$, the electronic dephasing rate $\gamma_{\rm el}=\Gamma_{\rm el}/2\pi$, the bosonic and vibrational mode frequencies $\nu_{\rm B}$ and $\nu_{\rm V}$, the relative coupling amplitude $\mu_{\rm V}/\mu_{\rm B}$, and the cross-manifold coupling $\mu_{\rm VB}$, yielding six free parameters per composition. 
The full pathways-resolved expressions and their derivation are given in the Supplementary Material.


\section*{Acknowledgments}
\paragraph*{Funding:}
This work was supported by University of Bordeaux for funding the project
\textit{"Nonlinear Spectroscopies of Aqueous Interfaces: study and structuration by intense terahertz pulses"} in the frame of the 2022 Pre-recruitment Chair Program.
\paragraph*{Author contributions:}
\textbf{Conceptualization:} MHK, JD, and EF.
\textbf{Data Curation:} MHK.
\textbf{Formal Analysis:} MHK (lead), JD, and EF.
\textbf{Funding Acquisition:} LD.
\textbf{Investigation:} MHK (lead) and TG.
\textbf{Methodology:} MHK, JD, and EF.
\textbf{Project Administration:} LD.
\textbf{Resources:} JD, TG, EA, LD, and EF.
\textbf{Software:} MHK.
\textbf{Supervision:} JD, LD, and EF.
\textbf{Validation:} MHK.
\textbf{Visualization:} MHK.
\textbf{Writing – Original Draft Preparation:} MHK (lead), JD, and EF.
\textbf{Writing – Review \& Editing:} MHK, JD, TG, EA, LD, and EF.
\paragraph*{Competing interests:}
There are no competing interests to declare.
\paragraph*{Data and materials availability:}
All data needed to evaluate the conclusions in the paper are present in the paper and/or the Supplementary Materials.


\bibliographystyle{elsarticle-num} 
\bibliography{THz-FWM_glass_surface}

\clearpage
\includepdf[pages=-]{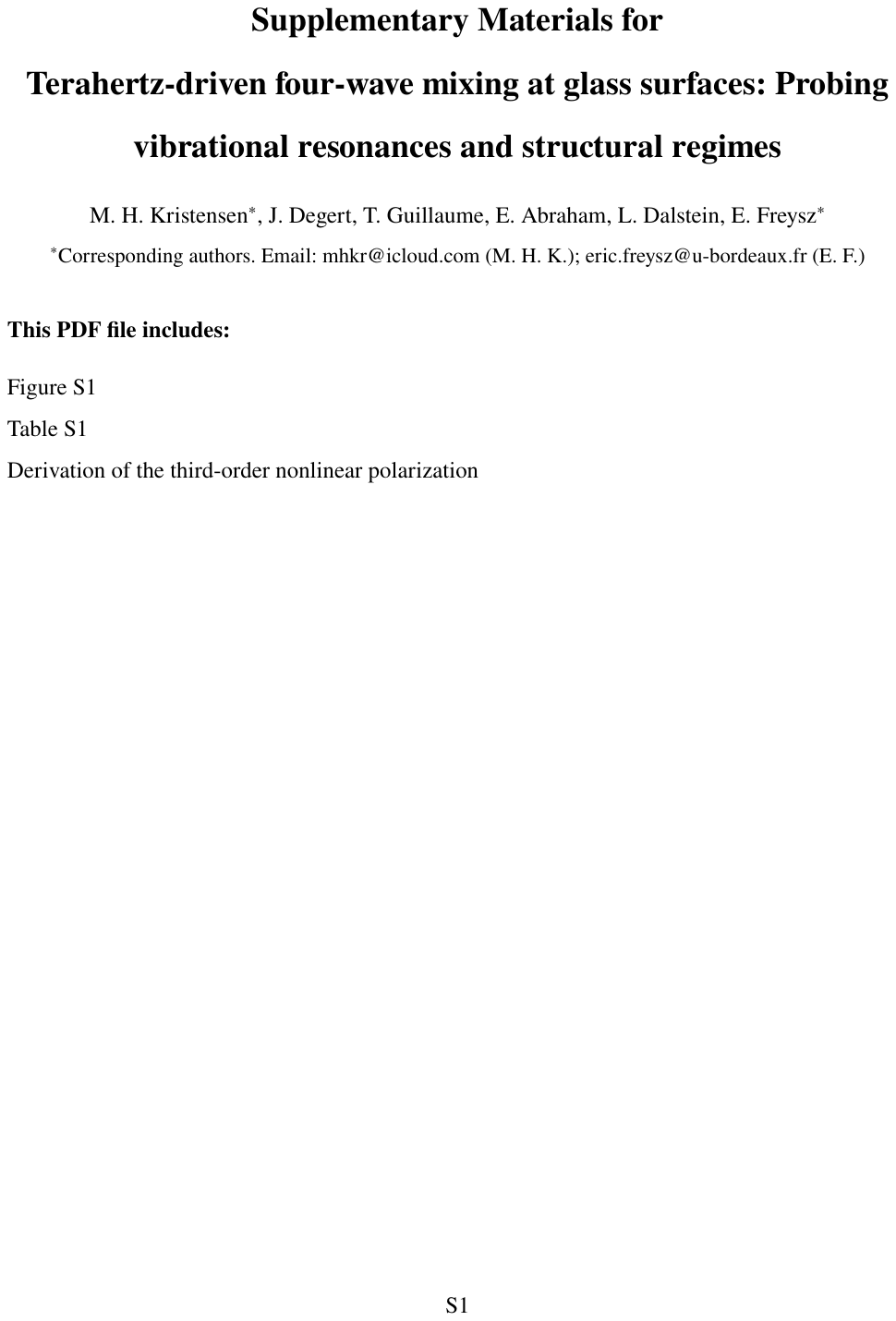} 

\end{document}